\documentclass[12pt]{article}
\usepackage{amssymb}

\begin{document}

\begin{center}
\textbf{SL(2, R)-SYMMETRY AND NONCOMMUTATIVE}

\smallskip\ 

\textbf{PHASE SPACE IN (2+2) DIMENSIONS}

\smallskip\ 

\smallskip\ 

J. A. Nieto \footnote{%
nieto@uas.uasnet.mx}

\smallskip\ 

\textit{Facultad de Ciencias F\'{\i}sico-Matem\'{a}ticas de la Universidad
Aut\'{o}noma} \textit{de Sinaloa, 80010, Culiac\'{a}n Sinaloa, M\'{e}xico.}

\bigskip\ 

\bigskip\ 

\textbf{Abstract}
\end{center}

We generalize the connection between 2t physics and noncommutative geometry.
In particular, we apply our formalism to a target spacetime of signature
(2+2). Specifically, we compute an algebra of a generalized SL(2,
R)-Hamiltonian constraint, showing that it satisfies a kind of algebra
associated with the noncommutative group $U_{\star }(1,1)$. We also comment
about a possible connection between our formalism and nonsymmetric
gravitational theory.

\bigskip\ 

\bigskip\ 

\bigskip\ 

\bigskip\ 

\bigskip\ 

\bigskip\ 

\bigskip\ 

\bigskip\ 


Pacs numbers: 04.60.-m, 04.65.+e, 11.15.-q, 11.30.Ly

April, 2009

\newpage

It is known that noncommutative field theory of 2t physics [1-2] relies on a
fundamental gauge symmetry principle based on the noncommutative group $%
U_{\star }(1,1)$ [3]. This approach originates from the observation that a
world-line theory admits a Lie algebra $sl_{\star }(2,R)$ gauge symmetry
acting on phase space $(q,p)$ [4]. Local considerations of the general
canonical transformations lead to the embedding of the corresponding
noncommutative algebra $sl_{\star }(2,R)$ into a bigger 4 parameter algebra $%
u_{\star }(1,1)$. However, it turns out that this noncommutative phase space
symmetry is based on the usual noncommutative relation between $q$ and $p$
rather than on the full noncommutative phase space that includes the
noncommutative configuration space of the $q$'s (and $p$'s) themselves. In
this work we prove that it makes sense to consider this more general
noncommutative phase space in 2t physics. We focus on the signature $2+2$
for at least two physical reasons; (1) it is the minimal possibility in 2t
physics and (2) it is an exceptional signature [5]. In principle, however,
our calculations also correspond to the more general case of $2+d$
dimensions.

Let us start recalling the traditional transition from a classical to a
quantum mechanical system. One may begin with the action

\begin{equation}
S[q]=\int_{t_{i}}^{t_{f}}dtL(q,\dot{q}),  \label{1}
\end{equation}%
where the Lagrangian $L$ is a function of the $q^{i}$-coordinates and the
corresponding velocities $\dot{q}^{i}\equiv dq^{i}/dt,$ with $i,j=1,\dots ,n$%
. One then defines the canonical momentum $p_{i}$ conjugate to $q^{i}$ as
follows;

\begin{equation}
p_{i}\equiv \frac{\partial L}{\partial \dot{q}^{i}},  \label{2}
\end{equation}%
and rewrites the action in the form

\begin{equation}
S[q,p]=\int_{t_{i}}^{t_{f}}dt(\dot{q}^{i}p_{i}-H(q,p)),  
\label{3}
\end{equation}%
where $H=H(q,p)$ is the canonical Hamiltonian,

\begin{equation}
H(q,p)\equiv \dot{q}^{i}p_{i}-L.  \label{4}
\end{equation}%
The transition to quantum mechanics is made by promoting the Hamiltonian $H$
as an operator $\hat{H}$ via the nonvanishing commutator

\begin{equation}
\lbrack \hat{q},\hat{p}]=-i,  \label{5}
\end{equation}%
with $\hbar =1$, and by writing the quantum formula

\begin{equation}
\hat{H}\mid \Psi >=0,  \label{6}
\end{equation}%
which determines the physical states $\mid \Psi >$ (see Refs. [6]-[8] for
details). Here, $[\hat{A},\hat{B}]=\hat{A}\hat{B}-\hat{B}\hat{A}$ denotes a
commutator for any arbitrary operators $\hat{A}$ and $\hat{B}$.

Recently, a new possibility to analyze the above program has emerged [9].
The key point for this new approach is the realization that since in the
action (3) there is a hidden invariance $SL(2,R)\approx Sp(2,R)\approx
SU(1,1)$ one may work in a unified canonical phase space of coordinates and
momenta. Let us recall how this hidden invariance emerges [4]. Consider
first the change of notation%
\begin{equation}
q_{1}^{i}\equiv q^{i},  \label{7}
\end{equation}%
and

\begin{equation}
q_{2}^{i}\equiv p^{i}.  \label{8}
\end{equation}%
Thus, by introducing the object $q_{a}^{i}$ with $a=1,2$ we see that these
two formulae can be unified. The next step is to rewrite (3) in terms of $%
q_{a}^{i}$ rather than in terms of $q^{i}$ and $p^{i}$. One find that up to
a total derivative the action (3) becomes [4] (see also Ref. [9])

\begin{equation}
S=\int_{t_{i}}^{t_{f}}dt\left( \frac{1}{2}\varepsilon ^{ab}\dot{q}%
_{a}^{i}q_{b}^{j}\delta _{ij}-H(q_{a}^{i})\right) .  \label{9}
\end{equation}%
Here, the symbol $\delta _{ij}$ denotes a Kronecker delta and $\varepsilon
^{ab}=-\varepsilon ^{ba}$, with $\varepsilon ^{12}=1$, is the antisymmetric $%
SL(2,R)$ invariant density. From this expression one observes that while the 
$SL(2,R)$-symmetry is hidden in (3) now it is manifest in the first term of
(9). Thus, it is natural to require the same $SL(2,R)$-symmetry for the
Hamiltonian $H(q_{a}^{i})$.

Consider the usual Hamiltonian for a free nonrelativistic point particle

\begin{equation}
H=\frac{p^{i}p^{j}\delta _{ij}}{2m},  \label{10}
\end{equation}%
with $i=\{1,2,3\}$. According to the notation (7)-(8) this becomes

\begin{equation}
H=\frac{q_{2}^{i}q_{2}^{j}\delta _{ij}}{2m}.  \label{11}
\end{equation}%
This Hamiltonian is not, of course, $SL(2,R)$-invariant. Adding a potential $%
V(q)$ to $H$ does not modify this conclusion. Thus, a Hamiltonian of the
form $H=\frac{p^{i}p^{j}\delta _{ij}}{2m}+V(q)$ does not admit a $SL(2,R)$%
-invariant formulation. It turns out that the same conclusion can be
obtained when one considers the relativistic Hamiltonian constraint $%
H=p^{i}p_{i}+m^{2}=0$, where in this case $i$ runs from $0$ to $3$.

The simplest example of $SL(2,R)$-invariant Hamiltonian seems to be

\begin{equation}
H=\frac{1}{2}\lambda ^{ab}q_{a}^{i}q_{b}^{j}\eta _{ij},  \label{12}
\end{equation}%
which can be understood as the Hamiltonian associated with a relativistic
harmonic oscillator in phase space. Here, we assume that $\lambda
^{ab}=\lambda ^{ba}$ is a Lagrange multiplier and $\eta
_{ij}=diag(-1,-1,1,1) $. (Notice that we are considering the special case of 
$2+2$ signature. The reason for this is that the symmetry of $SL(2,R)$
requires necessarily two times and two times with two space coordinates
provide an exceptional signature [5]. However, most our calculations below
can be easily generalized to $2+d$ dimensions.) Indeed, the Hamiltonian (12)
is a total Hamiltonian according to the terminology of the Dirac's
constraint hamiltonian systems formalism [10] (see also Refs. [6]-[8]). Let
us write (12) in the form

\begin{equation}
H=\frac{1}{2}\lambda ^{ab}Q_{ab},  \label{13}
\end{equation}%
where

\begin{equation}
Q_{ab}=q_{a}^{i}q_{b}^{j}\eta _{ij}  \label{14}
\end{equation}%
can be identified as the constraint of the physical system. Observe that the
constraint $Q_{ab}\approx 0$ is symmetric in the indices $a$ and $b$, that
is, $Q_{ab}=Q_{ba}$. (Here the symbol "$\approx $" means weakly equal to
zero [6]-[8].) Moreover, $Q_{ab}$ is a first class constraint and this can
be verified as follows. First note that using the definitions (7) and (8) we
can write the usual Poisson bracket, for arbitrary functions $f(q.p)$ and $%
g(q,p)$ of the canonical variables $q$ and $p$,

\begin{equation}
\{f,g\}=\frac{\partial f}{\partial q^{i}}\frac{\partial g}{\partial p_{i}}-%
\frac{\partial f}{\partial p_{i}}\frac{\partial g}{\partial q^{i}},
\label{15}
\end{equation}%
in the form

\begin{equation}
\{f,g\}=\varepsilon _{ab}\eta ^{ij}\frac{\partial f}{\partial q_{a}^{i}}%
\frac{\partial g}{\partial q_{b}^{j}}.  \label{16}
\end{equation}%
Thus, from (16) one discovers that

\begin{equation}
\{q_{a}^{i},q_{b}^{j}\}=\varepsilon _{ab}\eta ^{ij}.  \label{17}
\end{equation}%
Now, using the formula (17) it is straightforward to compute $%
\{Q_{ab},Q_{cd}\}$. Explicitly, we find

\begin{equation}
\begin{array}{c}
\{Q_{ab},Q_{cd}\}=\{q_{a}^{i}q_{b}^{j}\eta _{ij},q_{c}^{k}q_{d}^{l}\eta
_{kl}\} \\ 
\\ 
=(\{q_{a}^{i},q_{c}^{k}\}q_{b}^{j}q_{d}^{l}+\{q_{a}^{i},q_{d}^{l}%
\}q_{b}^{j}q_{c}^{k}+\{q_{b}^{j},q_{c}^{k}\}q_{a}^{i}q_{d}^{l}+%
\{q_{b}^{j},q_{d}^{l}\}q_{a}^{i}q_{c}^{k})\eta _{ij}\eta _{kl} \\ 
\\ 
=(\varepsilon _{ac}\eta ^{ik}q_{b}^{j}q_{d}^{l}+\varepsilon _{ad}\eta
^{il}q_{b}^{j}q_{c}^{k}+\varepsilon _{bc}\eta
^{jk}q_{a}^{i}q_{d}^{l}+\varepsilon _{bd}\eta ^{jl}q_{a}^{i}q_{c}^{k})\eta
_{ij}\eta _{kl}.%
\end{array}
\label{18}
\end{equation}%
This implies the algebra,%
\begin{equation}
\{Q_{ab},Q_{cd}\}=\varepsilon _{ac}Q_{bd}+\varepsilon
_{ad}Q_{bc}+\varepsilon _{bc}Q_{ad}+\varepsilon _{bd}Q_{ac}.  \label{19}
\end{equation}%
Thus, since we are assuming $Q_{ab}\approx 0,$ one sees that $%
\{Q_{ab},Q_{cd}\}\approx 0$ which means that $Q_{ab}$ is a first class
constraint. It turns out that $Q_{ab}$ can be also understood as the gauge
generator of the $SL(2,R)$-symmetry which is in fact determined by the
algebra (19) (see Ref. [11])

At the quantum level we promote $Q_{ab}$ as an operator $\hat{Q}_{ab}$ and
write%
\begin{equation}
\lbrack \hat{Q}_{ab},\hat{Q}_{cd}]=-i(\varepsilon _{ac}\hat{Q}%
_{bd}+\varepsilon _{ad}\hat{Q}_{bc}+\varepsilon _{bc}\hat{Q}%
_{ad}+\varepsilon _{bd}\hat{Q}_{ac})  \label{20}
\end{equation}%
and

\begin{equation}
\hat{Q}_{ab}\mid \Psi >=0.  \label{21}
\end{equation}%
Explicitly, the nonvanishing brackets of the algebra (20) can decomposed in
the form

\begin{equation}
\lbrack \hat{Q}_{11},\hat{Q}_{22}]=-4i\hat{Q}_{12},  \label{22}
\end{equation}%
\begin{equation}
\lbrack \hat{Q}_{11},\hat{Q}_{12}]=-2i\hat{Q}_{11},  \label{23}
\end{equation}%
and%
\begin{equation}
\lbrack \hat{Q}_{22},\hat{Q}_{12}]=+2i\hat{Q}_{22}.  \label{24}
\end{equation}%
By writing $\hat{J}_{3}=-\frac{1}{2}\hat{Q}_{12}$, $\hat{J}_{1}=\frac{1}{4}(%
\hat{Q}_{11}+\hat{Q}_{22})$ and $\hat{J}_{2}=\frac{1}{4}(\hat{Q}_{11}-\hat{Q}%
_{22})$ one finds the result

\begin{equation}
\lbrack \hat{J}_{1},\hat{J}_{2}]=-i\hat{J}_{3}  \label{25}
\end{equation}%
\begin{equation}
\lbrack \hat{J}_{1},\hat{J}_{3}]=i\hat{J}_{2},  \label{26}
\end{equation}%
and%
\begin{equation}
\lbrack \hat{J}_{2},\hat{J}_{3}]=i\hat{J}_{1},  \label{27}
\end{equation}%
which can be obtained from the algebra

\begin{equation}
\lbrack \hat{J}_{AB},\hat{J}_{CD}]=i(\eta _{AC}\hat{J}_{BD}-\eta _{AD}\hat{J}%
_{BC}+\eta _{BD}\hat{J}_{AC}-\eta _{BC}\hat{J}_{AD}),  \label{28}
\end{equation}%
where\ $\eta _{AB}=(-1,1,1),$ $\hat{J}_{AB}=-\hat{J}_{BA}$ and $\hat{J}^{A}=%
\frac{1}{2}\epsilon ^{ABC}\hat{J}_{BC}$, with $\epsilon ^{123}=1$ and $%
\epsilon _{123}=-1$. This is one way to see that $SL(2,R)$ is in fact the
covering group of $SO(1,2)$.

We would like now to generalize (17) in the form

\begin{equation}
\{q_{a}^{i},q_{b}^{j}\}=\varepsilon _{ab}\eta ^{ij}+g_{ab}\Omega ^{ij}.
\label{29}
\end{equation}%
Here, $\Omega ^{ij}$ is skew-simplectic form which can be chosen as%
\begin{equation}
\Omega _{ij}=\left( 
\begin{array}{cccc}
0 & -1 & 0 & 0 \\ 
1 & 0 & 0 & 0 \\ 
0 & 0 & 0 & -1 \\ 
0 & 0 & 1 & 0%
\end{array}%
\right) .  \label{30}
\end{equation}%
By convenience in (29) we choose $\eta _{ij}=diag(-1,1-1,1)$ rather than $%
\eta _{ij}=diag(-1,-1,1,1)$ and $g_{ab}=diag(\theta ,\phi ).$ Here $\theta $
and $\phi $ are two constant parameters. Note that $\eta _{ij}$ corresponds
to a flat signature . The reason for this choice for $\eta _{ij}$, among
other things, is because using the signature $(1+1)+(1+1)$ instead of $(2+2)$
some calculations are simplified. It is important to mention that the
expression (30) differs from the usual simplectic structure $\left( 
\begin{array}{cc}
0 & \delta _{ij} \\ 
-\delta _{ij} & 0%
\end{array}%
\right) $ by a change of bases. So, one can also think on (30) as a
consequence of the Darboux theorem.

One can prove that%
\begin{equation}
\Omega ^{ij}=\left( 
\begin{array}{cccc}
0 & 1 & 0 & 0 \\ 
-1 & 0 & 0 & 0 \\ 
0 & 0 & 0 & 1 \\ 
0 & 0 & -1 & 0%
\end{array}%
\right) .  \label{31}
\end{equation}%
In particular, using (30) and (31) we find the result $\Omega ^{ij}\Omega
^{kl}\eta _{jl}=-\eta ^{ik}$.

The generalization (29) motivates us to modify also the Hamiltonian
constraint (13) as follows

\begin{equation}
\mathcal{H=}\frac{1}{2}\Lambda ^{ab}\Sigma _{ab},  \label{32}
\end{equation}%
where $\Lambda ^{ab}$ are new Lagrange multipliers no necessary symmetric in
the indices $a$ and $b$ and

\begin{equation}
\Sigma _{ab}=q_{a}^{i}q_{b}^{j}(\eta _{ij}+i\xi \Omega _{ij}).  \label{33}
\end{equation}%
Here, $\xi $ is another constant parameter. Of course, the constraint $%
\Sigma _{ab}$ reduces to $Q_{ab}$ when $\xi \rightarrow 0$. We have

\begin{equation}
\begin{array}{c}
\{\Sigma _{ab},\Sigma _{cd}\}=\{q_{a}^{i}q_{b}^{j}\gamma
_{ij},q_{c}^{k}q_{d}^{l}\gamma _{kl}\} \\ 
\\ 
=(\{q_{a}^{i},q_{c}^{k}\}q_{b}^{j}q_{d}^{l}+\{q_{a}^{i},q_{d}^{l}%
\}q_{b}^{j}q_{c}^{k}+\{q_{b}^{j},q_{c}^{k}\}q_{a}^{i}q_{d}^{l}+%
\{q_{b}^{j},q_{d}^{l}\}q_{a}^{i}q_{c}^{k})\gamma _{ij}\gamma _{kl} \\ 
\\ 
=((\varepsilon _{ac}\eta ^{ik}+g_{ac}\Omega
^{ik})q_{b}^{j}q_{d}^{l}+(\varepsilon _{ad}\eta ^{il}++g_{ad}\Omega
^{il})q_{b}^{j}q_{c}^{k} \\ 
\\ 
+(\varepsilon _{bc}\eta ^{jk}++g_{bc}\Omega
^{jk})q_{a}^{i}q_{d}^{l}+(\varepsilon _{bd}\eta ^{jl}+g_{bd}\Omega
^{jl})q_{a}^{i}q_{c}^{k})(\eta _{ij}+i\xi \Omega _{ij})(\eta _{kl}+i\xi
\Omega _{kl}),%
\end{array}
\label{34}
\end{equation}%
where we used the definition

\begin{equation}
\gamma _{ij}=\eta _{ij}+i\xi \Omega _{ij},  \label{35}
\end{equation}%
which can be understood as an Hermitian metric since $\gamma _{ij}^{\dagger
}=\gamma _{ij}$. After some straightforward algebra we get%
\begin{equation}
\begin{array}{c}
\{\Sigma _{ab},\Sigma _{cd}\}=(1+\xi ^{2})\varepsilon _{ac}M_{bd}+(1-\xi
^{2})\varepsilon _{ad}M_{bc}+(1-\xi ^{2})\varepsilon _{bc}M_{ad} \\ 
\\ 
+(1+\xi ^{2})\varepsilon _{bd}M_{ac}+2i\xi \varepsilon _{ad}S_{cb}+2i\xi
\varepsilon _{bc}S_{ad}+(1+\xi ^{2})g_{ac}S_{bd}+(1-\xi ^{2})g_{ad}S_{bc} \\ 
\\ 
+(1+\xi ^{2})g_{bd}S_{ac}+(1-\xi ^{2})g_{bc}S_{ad}-2i\xi g_{ad}M_{bc}+2i\xi
g_{bc}M_{ad}.%
\end{array}
\label{36}
\end{equation}%
Here, we define $M_{ab}=q_{a}^{i}q_{b}^{j}\eta _{ij}=M_{ba}$ and $%
S_{ab}=q_{a}^{i}q_{b}^{j}\Omega _{ij}=-S_{ba}$. In other words we have%
\begin{equation}
\Sigma _{ab}=M_{ab}+i\xi S_{ab}.  \label{37}
\end{equation}%
Since in two dimensions we can always write%
\begin{equation}
S_{ab}=\kappa \varepsilon _{ab},  \label{38}
\end{equation}%
with $\kappa =\frac{1}{2}\varepsilon ^{ab}q_{a}^{i}q_{b}^{j}\Omega _{ij},$
we find that (36) is simplified in the form

\begin{equation}
\begin{array}{c}
\{\Sigma _{ab},\Sigma _{cd}\}=(1+\xi ^{2})\varepsilon _{ac}M_{bd}+(1-\xi
^{2})\varepsilon _{ad}M_{bc}+(1-\xi ^{2})\varepsilon _{bc}M_{ad} \\ 
\\ 
+(1+\xi ^{2})\varepsilon _{bd}M_{ac}+(1+\xi ^{2})g_{ac}S_{bd}+(1-\xi
^{2})g_{ad}S_{bc} \\ 
\\ 
+(1-\xi ^{2})g_{bc}S_{ad}+(1+\xi ^{2})g_{bd}S_{ac}-2i\xi g_{ad}M_{bc}+2i\xi
g_{bc}M_{ad}.%
\end{array}
\label{39}
\end{equation}%
It is not difficult to see that if the parameter $\xi $ and the quantity $%
\kappa $ vanish then (39) is reduced to (19). Thus, the expression (39)
provides a generalization of the algebra $sl(2,R)$. In fact, (39) seems to
correspond to the algebra associated with the noncommutative group $U_{\star
}(1,1)$. One way to understand this conclusion it is by observing that using
(38) the expression (37) can be written as

\begin{equation}
\Sigma _{ab}=M_{ab}+i\omega _{0}\varepsilon _{ab},  \label{40}
\end{equation}%
with $\omega _{0}=\xi \kappa $. It turns out that according to reference [1]
these are precisely the algebra generators associated with the
noncommutative group $U_{\star }(1,1)$ (see Ref. [3] for details). However,
in the way the expression (39) it is written, it is not clear whether it is
a closed algebra. In order to clarify this point we shall use the property
that in two dimensions it is always possible to choose a basis such that

\begin{equation}
M_{ab}=\rho g_{ab},  \label{41}
\end{equation}%
where $\rho \neq 0$ is an arbitrary function of the coordinates $q_{a}^{i}.$
Using this choice for $M_{ab}$ one finds that the last two terms of (39)
vanish. Thus, one discovers that (39) can be reduced in the form

\begin{equation}
\{\Sigma _{ab},\Sigma _{cd}\}=(\rho +\kappa )[(1+\xi ^{2})(\varepsilon
_{ac}g_{bd}+\varepsilon _{bd}g_{ac})+(1-\xi ^{2})(\varepsilon
_{ad}g_{bc}+\varepsilon _{bc}g_{ad}).  \label{42}
\end{equation}%
Now, from (40) and (41) one sees that $\Sigma _{ab}=\rho g_{ab}+i\omega
_{0}\varepsilon _{ab}.$ This implies that (42) can be written as

\begin{equation}
\{\Sigma _{ab},\Sigma _{cd}\}=\frac{(\rho +\kappa )}{\rho }[(1+\xi
^{2})(\varepsilon _{ac}\Sigma _{bd}+\varepsilon _{bd}\Sigma _{ca})+(1-\xi
^{2})(\varepsilon _{ad}\Sigma _{bc}+\varepsilon _{bc}\Sigma _{da})
\label{43}
\end{equation}%
or

\begin{equation}
\{\Sigma _{ab},\Sigma _{cd}\}=C_{abcd}^{ef}\Sigma _{ef},  \label{44}
\end{equation}%
where

\begin{equation}
C_{abcd}^{ef}=\frac{(\rho +\kappa )}{\rho }[(1+\xi ^{2})(\varepsilon
_{ac}\delta _{b}^{e}\delta _{d}^{f}+\varepsilon _{bd}\delta _{c}^{e}\delta
_{a}^{f})+(1-\xi ^{2})(\varepsilon _{ad}\delta _{b}^{e}\delta
_{c}^{f}+\varepsilon _{bc}\delta _{d}^{e}\delta _{a}^{f})].  \label{45}
\end{equation}%
Thus, we have shown that the algebra (39) can be written in the closed form
(44), with $C_{abcd}^{ef}$ playing the role of the structure "constants". Of
course, since we are assuming $\Sigma _{ab}\approx 0,$ from (40) one sees
that $\{\Sigma _{ab},\Sigma _{cd}\}\approx 0$ which means that $\Sigma _{ab}$
is a first class constraint.

Let us make some final analysis. First, consider the most general
prescription of the canonical variables $q_{a}^{i}$,

\begin{equation}
\{q_{a}^{i},q_{b}^{j}\}=\theta _{ab}^{ij},  \label{46}
\end{equation}%
which can be obtained from the generalized bracket

\begin{equation}
\{f,g\}=\theta _{ab}^{ij}\frac{\partial f}{\partial q_{a}^{i}}\frac{\partial
g}{\partial q_{b}^{j}}.  \label{47}
\end{equation}%
Since we can always decompose any matrix $B^{ij}=B^{(ij)}+B^{[ij]}$ in its
symmetric $B^{(ij)}$ and antisymmetric $B^{[ij]}$ parts, one finds that (47)
can also be written as

\begin{equation}
\{f,g\}=(\theta _{(ab)}^{(ij)}+\theta _{\lbrack ab]}^{(ij)}+\theta
_{(ab)}^{[ij]}+\theta _{\lbrack ab]}^{[ij]})\frac{\partial f}{\partial
q_{a}^{i}}\frac{\partial g}{\partial q_{b}^{j}}.  \label{48}
\end{equation}%
It is not difficult to realize that if we want that the bracket $\{f,g\}$
determines a simplectic structure we must set $\theta _{(ab)}^{(ij)}=0$ and $%
\theta _{\lbrack ab]}^{[ij]}=0$. Therefore, (48) can be reduced to

\begin{equation}
\{f,g\}=(\theta _{\lbrack ab]}^{(ij)}+\theta _{(ab)}^{[ij]})\frac{\partial f%
}{\partial q_{a}^{i}}\frac{\partial g}{\partial q_{b}^{j}}.  \label{49}
\end{equation}%
This can be simplified further by considering that in two dimensions we can
always write $\theta _{\lbrack ab]}^{(ij)}=\varepsilon _{ab}\eta ^{ij}$
where we assumed a flat "spacetime" $\eta ^{ij}$ (Of course, in a more
general case one can assume a curved metric $g^{ij}$.) Similarly, in two
dimensions we can write $\theta _{(ab)}^{[ij]}=g_{ab}\Omega ^{ij}$, where $%
g_{ab}=g_{ba}$ is a two dimensional metric and $\Omega ^{ij}=-\Omega ^{ji}$.
In order to distinguish between $q$'s and $p$'s we choose a basis such that $%
g_{ab}=diag(\theta ,\phi )$, but in principle in two dimensions one can
always find a basis such that $g_{ab}\rightarrow \sigma \delta _{ab}$, where 
$\sigma $ is a constant conformal factor. Thus, we have proved that the most
general meaningful simplectic structure is provided by the bracket

\begin{equation}
\{f,g\}=(\varepsilon _{ab}\eta ^{ij}+g_{ab}\Omega ^{ij})\frac{\partial f}{%
\partial q_{a}^{i}}\frac{\partial g}{\partial q_{b}^{j}}.  \label{50}
\end{equation}%
It turns out that this expression leads precisely to our generalized bracket
(29). Notice that the above calculation is true for any even spacetime
dimension. In terms of $q^{i}$ and $p^{j}$ one finds that the algebra (29)
becomes

\begin{equation}
\{q^{i},p^{j}\}=\eta ^{ij},  \label{51}
\end{equation}

\begin{equation}
\{q^{i},q^{j}\}=\theta \Omega ^{ij},  \label{52}
\end{equation}%
and

\begin{equation}
\{p^{i},p^{j}\}=\phi \Omega ^{ij}.  \label{53}
\end{equation}%
(See Refs. [12] and [13] for an alternative construction.) We still need to
justify that the quantities $\theta $ and $\phi $ can be chosen as a
constant parameters. Let us first introduce new canonical variables $\tilde{q%
}_{a}^{i}$ such that

\begin{equation}
q_{ai}=a_{aij}^{b}\tilde{q}_{b}^{j},  \label{54}
\end{equation}%
where $q_{ai}=\eta _{ij}q_{a}^{j}$ and $a_{aij}^{b}=g_{ac}a_{ij}^{cb}$.
Writing $a_{ij}^{ab}=g_{ij}^{ab}+A_{ij}^{ab}$ with $g_{ij}^{ab}=\frac{1}{2}%
(a_{ij}^{ab}+a_{ij}^{ba})$ and $A_{ij}^{ab}=\frac{1}{2}%
(a_{ij}^{ab}-a_{ij}^{ba})$ we see that one can always write (54) as

\begin{equation}
q_{ai}=g_{aij}^{b}\tilde{q}_{b}^{j}+A_{aij}^{b}\tilde{q}_{b}^{j}.  \label{55}
\end{equation}%
We require that the new variables $\tilde{q}_{a}^{i}$ satisfy the usual
canonical algebra

\begin{equation}
\{\tilde{q}_{a}^{i},\tilde{q}_{b}^{j}\}=\varepsilon _{ab}\eta ^{ij},
\label{56}
\end{equation}%
or

\begin{equation}
\{\tilde{q}^{i},\tilde{p}^{j}\}=\eta ^{ij},  \label{57}
\end{equation}

\begin{equation}
\{\tilde{q}^{i},\tilde{q}^{j}\}=0,  \label{58}
\end{equation}%
and

\begin{equation}
\{\tilde{p}^{i},\tilde{p}^{j}\}=0,  \label{59}
\end{equation}%
where we used the corresponding definitions (7) and (8) for $\tilde{q}%
_{a}^{i}$. The authors of Ref. [14] have shown that a meaningful result can
be obtained if $g_{ij}^{ab}=g_{ji}^{ab}$ and $A_{ij}^{ab}=-A_{ji}^{ab}$,
that is, if $g_{ij}^{ab}$ is symmetric in both kind of indices $a,b$ and $i,$
$j$, and $A_{ij}^{ab}$ is antisymmetric in both kind of indices $a,b$ and $i$%
, $j$. Moreover, these authors show that one can assume $g_{1ij}^{1}=\alpha
\eta _{ij}$, $g_{2ij}^{2}=\beta \eta _{ij}$, while $%
g_{1ij}^{2}=g_{2ij}^{1}=0 $. Here, $\alpha $ and $\beta $ are, in principle,
two different constant parameters. In addition, one can take $%
A_{ij}^{ab}=\varsigma \varepsilon ^{ab}\Omega _{ij}$, with $\varsigma
=(\alpha \beta -1)^{1/2}$. Thus, these results can be summarized by writing
(55) in the form

\begin{equation}
q^{i}=\alpha \tilde{q}^{i}-\varsigma \Omega ^{ij}\tilde{p}_{j}  \label{60}
\end{equation}%
and

\begin{equation}
p^{i}=\beta \tilde{p}^{i}+\varsigma \Omega ^{ij}\tilde{q}_{j}.  \label{61}
\end{equation}%
Solving $\tilde{q}_{i}$ and $\tilde{p}_{i}$ in terms of $q^{i}$ and $p^{j}$
one finds [15]%
\begin{equation}
\tilde{q}^{i}=\frac{1}{\rho }(\beta q^{i}+\varsigma \Omega ^{ij}p_{j})
\label{62}
\end{equation}%
and

\begin{equation}
\tilde{p}^{i}=\frac{1}{\rho }(\alpha p^{i}-\varsigma \Omega ^{ij}q_{j}).
\label{63}
\end{equation}%
Here, $\rho =2\alpha \beta -1$.

Using (57)-(61) we obtain

\begin{equation}
\{q^{i},p^{j}\}=\eta ^{ij},  \label{64}
\end{equation}

\begin{equation}
\{q^{i},q^{j}\}=2\alpha \varsigma \Omega ^{ij}  \label{65}
\end{equation}%
and

\begin{equation}
\{p^{i},p^{j}\}=2\beta \varsigma \Omega ^{ij}.  \label{66}
\end{equation}%
By comparing (64)-(66) with (51)-(53) we see that one must set $\theta
=2\alpha \varsigma =2\alpha (\alpha \beta -1)^{1/2}$ and $\phi =2\beta
\varsigma =2\beta (\alpha \beta -1)^{1/2}$. Thus, one discovers that these
results not only prove that it makes sense to choose $\theta $ and $\phi $
as a constant parameters but also assure that due to (57)-(59) any three
arbitrary functions $f(\tilde{q}_{a}^{i})$, $g(\tilde{q}_{a}^{i})$ and $h(%
\tilde{q}_{a}^{i})$ satisfy automatically the Jacobi identity. In turn, this
implies that the variables $\Sigma _{ab}$ must also satisfy the Jacobi
identity.

Another possible consequence of our formalism is that we can develop
noncommutative field theory by defining the noncommutative Moyal start
product as follows

\begin{equation}
(F\star G)(q_{a}^{i})=\exp (\theta _{ab}^{ij}\frac{\partial }{\partial
q_{a}^{i}}\frac{\partial }{\partial \tilde{q}_{b}^{j}})F(q_{a}^{i})G(\tilde{q%
}_{a}^{i})\mid _{q_{a}^{i}=\tilde{q}_{a}^{i}},  \label{67}
\end{equation}%
with $\theta _{ab}^{ij}=\varepsilon _{ab}\eta ^{ij}+g_{ab}\Omega ^{ij}$.
Consequently, one can define the star commutator between any two field $%
F(q_{a}^{i})$ and $G(q_{a}^{i})$ as

\begin{equation}
\lbrack F,G]_{\star }=F\star G-G\star F.  \label{68}
\end{equation}%
In particular, it may be interesting for further research to apply this
construction to the cases of gravity [16]--[18], self-dual gravity [19] and
area-preserving diffeomorphisms in gauge theory on a non-commutative plane
[20].

It seems also interesting to generalize the constraint Hamiltonian (32) to a
curved spacetime as follows

\begin{equation}
\mathcal{H}=\frac{1}{2}\Lambda
^{ab}q_{b}^{i}q_{b}^{j}(g_{ij}(q_{c}^{k})+iA_{ij}(q_{c}^{k})).  \label{69}
\end{equation}%
Here $g_{ij}(q_{c}^{k})=g_{ji}(q_{c}^{k})$ is a curved spacetime metric and $%
A_{ij}=-A_{ji}$ is antisymmetric gauge field. This generalizes the metric
(35) in the form%
\begin{equation}
\varphi _{ij}(q_{c}^{k})=g_{ij}(q_{c}^{k})+iA_{ij}(q_{c}^{k}),  \label{70}
\end{equation}%
which is also a Hermitian metric. It turns out that this kind of metric is
the main mathematical object in nonsymmetric gravitational theories (see
Refs. [21]-[23] and references therein). But of course our metric refers to
the phase space rather to the configuration space itself. At this respect it
is worth mentioning that in Ref. [24] it is provided evidence for a position
and momentum dependent metric in 2t physics.

Moreover, it turns out interesting to write (70) in the alternative
vielbeins form

\begin{equation}
\varphi _{ij}(q_{c}^{k})=e_{i}^{(m)}(q_{c}^{k})e_{j}^{(n)}(q_{c}^{k})\eta
_{(mn)}+if_{i}^{(m)}(q_{c}^{k})f_{j}^{(n)}(q_{c}^{k})\Omega _{(mn)},
\label{71}
\end{equation}%
with $g_{ij}=e_{i}^{(m)}(q_{c}^{k})e_{j}^{(n)}(q_{c}^{k})\eta _{(mn)}$ and $%
A_{ij}=f_{i}^{(m)}(q_{c}^{k})f_{j}^{(n)}(q_{c}^{k})\Omega _{(mn)}$. This way
to write (70) it suggests to consider a star product deformation $%
E_{i}^{(m)}\star E_{i}^{(m)}$, where we have introduced the complex vielbien
field $E_{i}^{(m)}=e_{i}^{(m)}+if_{i}^{(m)}$. This should lead of course to
an infinite number of corrections to (70). Moreover, we should mention that
the Moyal product in curved phase space [25] has already been studied by a
number of authors, including Fedosov [26] and Kontsevich [27] (see also Ref.
[28]). However, it seems that the particular case of 2+2 dimensions has not
been considered. In any case the metric (70) seems to determine a bridge
between our formalism and nonsymmetric gravitational theory, which we expect
to explore in more detail in the coming future.

The present work it might be also relevant in connection with the Ref. [29]
where there is a region with two-times in $U_{\star }(1,1)\times U_{\star
}(1,1)$ noncommutative gauge theory formulation of 3D gravity.

It has been established [30] a connection between 2t physics and oriented
matroid theory [31] (for a connection between oriented matroid theory and
other scenarios in high energy physics see Refs. [32]-[33] and references
therein). Since the $2+2$ signature is linked to brane physics [34] which in
turn it is connected to oriented matroids [35-36] it may also be physically
interesting for further directions to investigate the relation between the
present formalism and all these scenarios in the context of oriented matroid
theory.

We should mention a number of interesting topics that may be related to the
present formalism. It is known that in the Yang's algebra the coordinates
and the momenta are also not commuting [37]-[39]. The relevant group in this
case is the conformal group $SO(2,4)$ which has also an important status in
2t physics (see Refs. [1], [2] and [4]). Another direction for extensions of
our calculations is the possibility to include in the discussion the quantum
group concept. At this respect the work by Majid [40] may be of special help
since as this author emphasize "Lie groups are the simplest Riemann
manifolds and quantum groups are the simplest noncommutative spaces". So,
quantum groups are deeply connected with noncommutative geometry and in this
direction the Refs. [41] and [42] may be specially useful. Finally,
bi-Hamiltonian structure for integrable models (see Refs [43]-[45] and
references there in) may require also two times and one wonders whether our
formalism may also find an important application in such a subject.

\bigskip\ 

\begin{center}
\textbf{Acknowledgments}
\end{center}

I would like to thank the second referee for emphasizing the possible
importance of the formulae (71) and for valuable comments concerning new
directions of development of the present work. I also appreciate the helpful
comments of M. Sabido, O. Velarde and I. Le\'{o}n-Monz\'{o}n. This work was
partially supported by PROFAPI 2007 and PIFI 3.3.

\bigskip\

\end{document}